\documentclass[12pt]{article}
\usepackage{graphicx,abstract,epstopdf,times,authblk,geometry,fancyhdr,etoolbox,caption,sectsty,float}
\usepackage[sort,round]{natbib}

\newgeometry{top=2cm,left=2.5cm,right=2.5cm,bottom=4.5cm}

\date{\vspace{-\baselineskip}\vspace{-\baselineskip}}
\allsectionsfont{\fontsize{12pt}{14pt}\selectfont\bfseries }

\AtBeginEnvironment{tabular}{\fontsize{10pt}{9pt}\selectfont}
\newcommand\emails{\affil[ ]}
\usepackage{siunitx}

\title{Liquid Humans - Pedestrian Simulator based on the LWR-model}

\author[1]{Quirin Aumann}
\author[1]{Carlos M. Osorio}
\author[1]{Celeste Lai}   

\emails{quirin.aumann@tum.de , carlos.osorio@tum.de, celeste.lai@tum.de}

\affil[1]{Chair of Computational Modeling and Simulation, Technische Universit\"at M\"unchen}

\begin{document}

\maketitle

\begin{abstract}
Dense human flow has been a concern for the safety of public events for a long time. Macroscopic pedestrian models, which are mainly based on fluid dynamics, are often used to simulate huge crowds due to their low computational costs \citep{Colombo2005}. Similar approaches are used in the field of traffic simulations \citep{Lighthill1955}. A combined macroscopic simulation of vehicles and pedestrians is extremely helpful for all-encompassing traffic control. Therefore, we developed a hybrid model that contains networks for vehicular traffic and human flow. This comprehensive model supports concurrent multi-modal simulations of traffic and pedestrians \citep{Paper}.
\end{abstract}

\section{Introduction}


\subsection{Hartmann-Sivers Model}
Every year, about 2000 people die due to the movement of large human crowds \citep{Hughes2003}. The growing eventisation of social and public life in the 21st century \citep{Hitzler2010} increases the probability and frequency of organized events attended by thousands or millions of people. In the past, fatal accidents have occurred repeatedly at such large events. Examples include the catastrophe in the Heysel stadium in 1985 \citep{Lewis1989} with 39 fatalities or the Loveparade disaster in 2010 with 21 deaths \citep{Hitzler2011}. Professional crowd control increases the safety of public events and lowers the probability of fatal  accidents in the context of large human flows. Pedestrian dynamics simulations support successful crowd management by helping organizers of public events to foresee and prevent dangerous situations. Macroscopic pedestrian dynamic models like the Hartmann \& Sivers mode \citep{Hartmann2013} are most suitable for simulating large numbers of people \citep{BiedermannFBI}. The same macroscopic model can be applied to simulate the motion of cars. By combining human and vehicular flow into one model, we can achieve an all-embracing simulation model, which covers crowd and traffic control at the same time \citep{Paper}. Due to the macroscopic nature of these simulation models, we can simulate very large and dense scenarios faster than real time.

\subsection{Current Hybrid Modeling}
Two different kinds of hybrid modeling exist in the field of traffic and crowd control. The first type combines traffic or pedestrian dynamcis models at different scales in order to reduce the overall computational effort. This means that simulations in the hazardous areas of the scenario (places with high densities or small bottlenecks) are calculated with high detail but computationally costly models. At the same time, less hazardous areas of the scenario can be simulated by less detailed but more cost-effective models. This provides us precious results in the interesting parts of the scenario with a reduced total computational cost. \citealp{Tolba2005} as well as \citealp{Mccrea2010} developed such hybrid models in the field of vehicular flow. Similar models were developed for pedestrian dynamics. A wide range of hybrid modeling exists; from the coupling of very specific pedestrian simulation models  (e.g. \citealp{Nguyen2012}) to generic coupling frameworks \citep{BiedermannTransiTUM}. An overview of the current state of the art can be found by \citealp{Ijaz2015}.

The second type of hybrid modeling does not combine different scales, but different kinds of simulations. \citealp{Galea2008} describe the coupling of pedestrian dynamics and fire simulations and \citealp{Goettlich2011} combine an evacuation simulation with the spread of hazardous gases. Our approach is the connection of vehicular traffic with human flow. Some studies have already been carried out in this field. \citealp{Pretto2011} coupled agent-based representations of pedestrians and cars. A macroscopic approach was developed by \citealp{Borsche2014}. They used the classic Lighthill-Whitham-Richards model \citep{Lighthill1955,Richards1956} for the vehicular flow and the model from \citealp{Hughes2002} for the macroscopic simulation of pedestrians. Their work is important progress in the field of hybrid macroscopic modeling. However, they considered vehicles and humans as two interacting, but separate flows. In reality, cars and pedestrians are not completely separate: drivers exit their cars after finishing driving and become pedestrians and pedestrians may enter their vehicles to become drivers. Our approach is a hybrid model, which is capable of converting humans and vehicles into each other to create a realistic simulation model.

\section{The Hybrid Simulation Model CarPed}

\subsection{The Hybrid approach}
\begin{figure}[!ht]
	\begin{center}
	\includegraphics[width=140mm]{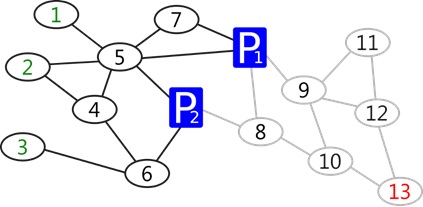}
	\caption{An Exemplary Network Consisting of Nodes and Edges}
	\label{AnExemplaryNetworkConsistingofNodesandEdges}
	\end{center}
\end{figure}

It is important to use a single network that combines pedestrian and traffic simulations. Therefore, we introduce an interface which links human and vehicular flow. The movement behavior of cars and pedestrians is based on the macroscopic Hartmann and Sivers model \citep{Hartmann2013} and is calculated on a discrete network. Roads for cars and walking paths for pedestrians are represented by the edges of this network. Its nodes represent the links between the different edges and can be considered crossings. Specifically, the nodes have different flow rates. There are special nodes, which represent parking lots, and serve as a connection between roads and walkways. Visitors of the event enter and leave their cars there. To put it another way, the simulated subjects get transformed from vehicles into pedestrians and vice versa. Therefore, the parking lots serve as a connection of the traffic and pedestrian networks. In Figure \ref{AnExemplaryNetworkConsistingofNodesandEdges}, an exemplary network is shown. The darker nodes and edges represent streets and their respective intersections, whereas the pedestrians’ walkways and intersections are represented in a lighter gray. The transformation of cars and pedestrians is carried out at special parking lot nodes. Green numbered entry nodes represent sources, which are the starting points of the visitors (e.g. their hometown). The red marked exit node is the final destination of the visitors (e.g. a large public event), where they are deleted from the network.

We implemented the Hartmann and Sivers model for pedestrian simulation and extended it to traffic simulation by using different model parameters. The extension was done by applying different maximum capacities to the edges and nodes, using different free flow velocities and other model constants for the simulated subjects. We used data obtained from \citealp{Weidmann1993} for the pedestrians and data from \citealp{Wachs2000} for the vehicle simulation. Additionally, we conducted a large field study to receive sufficient data for the transformation process (see Section 3).

\subsection{The Hartmann-and-Sivers Model}
The Hartmann and Sivers model adopts a structured continuum model on a macroscopic level. The approach basically relies on fundamental diagrams - the relation between fluxes and local densities - as well as the explicit consideration of individual velocities; thus showing better agreement with microscopic pedestrian models \citep{Hartmann2013}. The free flow velocity $v_{ff}$ is an individual property of each pedestrian / vehicle, i.e. a macroscopic, time independent state variable attached to each individual. The free flow velocity describes the highest velocity an individual can reach under optimal circumstances. This property is distributed normally among pedestrians or cars. Thus for modelling human or vehicular flow, the following equation holds
\begin{equation}
\frac{\partial{\rho(v_{ff};x,t)}}{\partial{t}} + \frac{\partial}{\partial{x}} (v\:(v_{ff};\rho(x,t))\rho(v_{ff};x,t)) = f(v_{ff};x,t)
\label{ContinuousMacroscopicModel}
\end{equation}

The total density of subjects is given by the sum of densities for all free flow velocities.
\begin{equation}
\rho(x,t) = \int_0^{v_{ff}^{max}} \rho(v;x,t)\mathrm{d}v
\label{TotalDensity}
\end{equation}

The fundamental relation between velocity and density for pedestrians can be adopted according to:
\begin{equation}
v(v_{ff};\rho)=v_{ff}\left[1-exp\left(-\gamma\left(\frac{1}{\rho}-\frac{1}{\rho_{max}}\right)\right)\right]
\label{RelationBetweenVelocityAndDensity}
\end{equation}

and for cars:
\begin{equation}
v(v_{ff};\rho) = v_{ff} \frac{\rho_{max}^n - \rho^n}{\rho_{max}^n + K\rho^n}
\label{RelationBetweenVelocityAndDensityCars}
\end{equation}

The maximum density $\rho_{max}$ and the factor $\gamma$ are unique parameters for pedestrians, as well as $n$ and $K$ for cars. The differential equations for pedestrians and cars were discretized by an upwind finite volume scheme \citep{Hartmann2013}.

\subsection{Discretization}

The used parameters are $\rho_{max}=5.4  \si{people/m^2}$, $\gamma=1.913$, $v_{ff}=1.34 \si{\metre/\second}$ in case of pedestrians, and  $\rho_{max}=0.12 \si{cars/m}$, $K=6.83$, $n=1.81$,  $v_{ff}=15$ m/s for cars in the city. The following finite difference is used:

\begin{equation}
\rho_{i}^{j+1}=\rho_{i}^{j}-\Delta t\left[d^{+}F_{\Delta}^{+}+d^{-}F_{\Delta}^{-}\right]
\end{equation}

In this case the proper \citealp{Hartmann2013} method is used:

\begin{equation}
F_{\Delta}^{+}=\frac{\rho_{i}v(v_{ff};(1-\alpha)\rho_{i}+\alpha\rho_{i+1})-\rho_{i-1}v(v_{ff};(1-\alpha)\rho_{i-1}+\alpha\rho_{i})}{\Delta x}
\end{equation}

\begin{equation}
F_{\Delta}^{-}=\frac{\rho_{i+1}v(v_{ff};(1-\alpha)\rho_{i+1}+\alpha\rho_{i})-\rho_{i}v(v_{ff};(1-\alpha)\rho_{i}+\alpha\rho_{i-1})}{\Delta x}
\end{equation}

Being $\alpha$ a discretization parameter between 0 and 1,  $d^{+}=1$  and $d^{-}=0$ always for forward moving cars and pedestrians. When the velocity is constant, this scheme reduces to first order upwind method.

\subsubsection{Simulation of structured model}

At first, the initial densities in the starting nodes separate into different velocity classes. If a very big event in which the number of pedestrians is very high (central limit theorem) is assumed, a normal distribution to predict how many people has each free flow velocity class can be used, using $N(1.34,0.26)$ as specified in \citealp{Hartmann2013}.

\begin{equation}
\rho_{j,0}=\rho_{0}\left[N.cumulativeProbability(v_{j})-N.cumulativeProbability(v_{j-1})\right]
\end{equation}

In this case, some 'lines' of people that walk are assumed, each of them with a different free flow velocity. A person never changes his or her own free flow velocity. Assuming that, the velocity can be further decomposed as:

\begin{equation}
v(v_{ff};\rho(x,t))=v_{ff}\hat{v}(\rho(x,t))
\end{equation}

The $\hat{v}$ part is dependant on the total density, and the $v_{ff}$ part is the well known normal-distributed free flow velocity for each class.

\subsection{Mass conservation}
The upwind schemes are in principle mass conservative within an infinite domain. That means as well as the values do not 'escape' through the boundaries the mass is conserved. However, the edges in this case are indeed finite so a scheme that takes people close to the boundaries before they leave and move them into other edges is necessary. 

The spatial discretization of an edge has $N$ cells, so the array goes from $0$ to $N-1$. The cells $0$ and $N-1$ are boundary conditions and do not belong to the solver domain, and their value is zero. In order to assure the mass conservation, the cell 1 (second cell) is considered the start cell so the people from the beginning node is placed there. The cell $N-2$ (penultimate cell) is considered the end cell of an edge and those people are sent to the end node when they reach the cell. The penultimate cell $N-2$ is then emptied there after. 

When some people reach the penultimate cell of an incoming edge, those people are sent to the node. Then, when the node is computed, the remaining people are placed in the outcoming edges according to the weights. The way to split the people, and assuming not-infinite nodes that let the people accumulate in the incoming edges are aspects that should be taken into account. Integrating the densities in the edges, and then summing the people in the nodes can be useful if the user wants to check if people are preserved through the computation.
\begin{equation}
people=PPC\sum_{s}w_{s}\Delta x\sum_{i=0}^{N_s-1}\rho_{s,i}+\sum_{w}w_{w}\Delta x\sum_{i=0}^{N_w-1}\rho_{w,i}+\sum_{n}\rho_{n}
\end{equation}

Where $PPC$ means persons per car, $s$ streets, $w$  walkways and $n$ nodes. In the nodes we calculate with the number of pedestrians and in the edges with the local density which is given in persons per square metre. Initializing people and cars in the start nodes is the first step in the computation.

\subsection{Fixed distributor}
In this distributor, the same density is placed for all the outcoming edges so the number of people that an edge takes is proportional to the width. This density can be found using the following equation.
\begin{equation}
\sum\rho_{in}w_{in}dx=\sum\rho_{out}w_{out}dx
\end{equation}

\begin{equation}
\rho_{out}=\frac{\sum\rho_{in}w_{in}dx}{\sum w_{out}dx}
\end{equation}

In the start nodes the entering people and cars are initialized using the information from the GUI. Sometimes the $\rho_{out}$ is greater than the maximum density allowed in the edges. This is a situation that happens quite often i.e. in a sudden change of width in two consecutive edges. If the first edge is very wide, a lot of people will want to go into the following edge (in this case a very narrow one). The first solution is assuming infinite nodes, so in that case the people will wait until the second cell of the outcoming edge allows them.

\subsection{Finite nodes}
In situations like the one explained before, infinite nodes shall not be used because such behavior does not happen in reality. The people should go inside the node and then leave immediately, and if they find the end node full, they should stay in the incoming edge instead of getting inside that full node. Unfortunately, the first order equation and the structured model leads the fast people go first and always the density is a decreasing function through the edge. The model equation by itself is far away from concentrating the people close to the edge exit when the node is full. Also, the parameter $\alpha=1$ (remember Hartmann and Sivers numerical method) is more 'human-wise' because the people can 'look forward' and use that density to calculate their velocity. This model and numerical method does not look beyond one cell density in the edge. Implementing a numerical method or another model with longer sight for the velocity, could make the people move
in a more human-wise way.

\begin{figure}[!ht]
	\begin{center}
	\includegraphics[width=140mm]{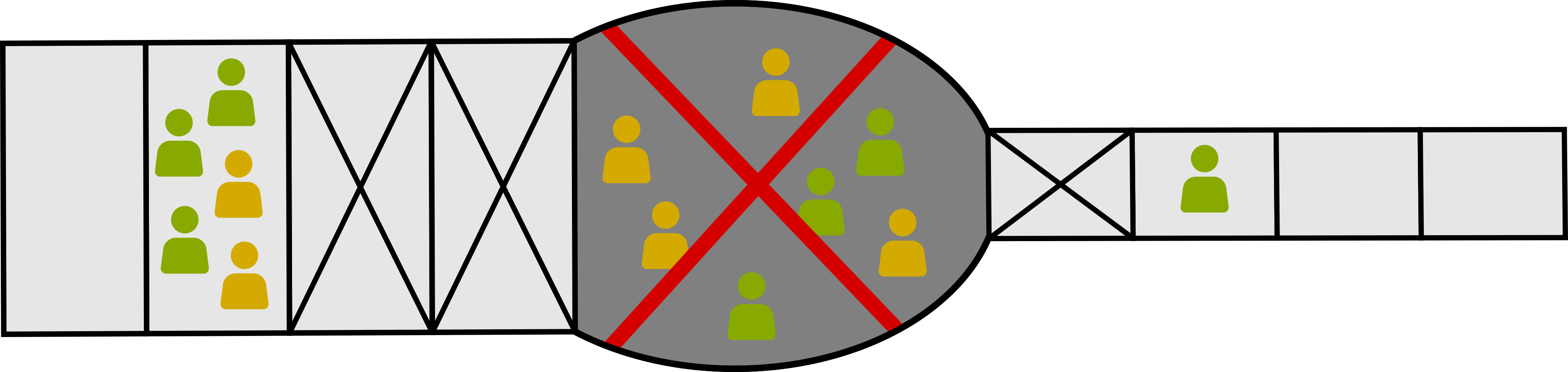}
	\caption{An Example Showing a Full Cell}
	\label{AnExampleShowingaFullCell}
	\end{center}
\end{figure}

In order to  allow the people to accumulate before a full node (see Figure \ref{AnExampleShowingaFullCell}) in the incoming edge, some new routines must be implemented. Section \ref{section:code} explains these algorithms in detail.

Figure \ref{Narrowing_results} shows the behavior of pedestrians in a narrowing example, in both infinite and finite nodes:
\begin{figure}[!ht]
	\centering%
	\begin{tabular}{cc}
	\multicolumn{1}{c}{\includegraphics[scale=0.6]{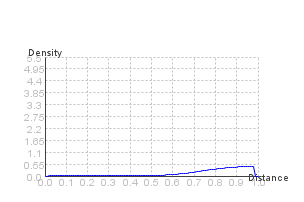}} & \includegraphics[scale=0.6]{{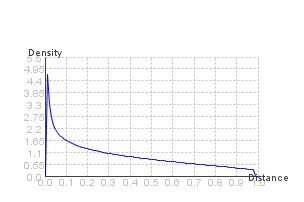}}\tabularnewline
	a) Edge 1 (incoming) infinite node & b) Edge 2 (outcoming) infinite node\tabularnewline
	\includegraphics[scale=0.6]{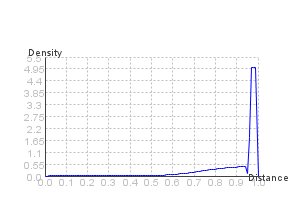} & \includegraphics[scale=0.6]{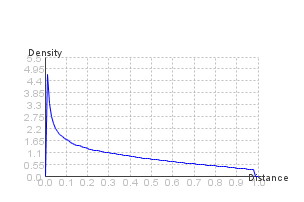}\tabularnewline
	c) Edge 1 (incoming) finite node & d) Edge 2 (outcoming) finite node\tabularnewline
	\end{tabular}
	\caption{Narrowing example ratio 30:1. 500 timesteps. $\alpha=1,\ dx=0.01,\ dt=0.002,\ L_{1}=1,\ L_{2}=1,\ w_{1}=30,\ w_{2}=1$.}
	\label{Narrowing_results}
\end{figure}

\subsection{Routing Behavior of Pedestrian and Cars}
In their original paper, Hartmann and Sivers used fixed probability values to distribute the pedestrians at each intersection. These values were obtained by comparing the capacities of the adjacent edges and did not take the current density of those edges into account. It is possible that the majority of a flow gets directed into an edge which is already relatively full, and the required time for the flow to pass the system could increase in an unrealistic
way. Therefore, we extend this approach and introduce a routing algorithm which determines the possible edges the pedestrians or cars will be allocated at.\\\\
To find the fastest path through the system, we use the classical Dijkstra’s algorithm \citep{Dijkstra1959}. This algorithm weighs all edges of the graph to compute the shortest way through the system. We used the Dijkstra’s algorithm to calculate the route with the shortest travel time. Therefore, we used the length and the current velocity $v(v_{ff};\rho)$ as weighting factors:

\begin{equation}
Weight = \frac{length}{v(v_{ff};\rho)}
\label{Eq:Four}
\end{equation}
\\
With increasing density the velocity $v(v_{ff};\rho)$ decreases according to Equation \ref{RelationBetweenVelocityAndDensity}. If the density of an edge reaches the maximum density $\rho_{max}$, the edge is considered as closed. Closed edges are ignored for the calculation of the routing algorithm.

\subsection{Transformation Between The Two Simulation Models}
Cars get transformed into pedestrians in the parking lot nodes. Typically, for ordinary cars, one to five people get out of each vehicle, with a prescribed random distribution. The distribution for the transformation was gathered from field data \citep{Paper}. The transformed pedestrians and cars receive their individual parameters, like their free flow velocity or their maximum density, according to \citealp{Weidmann1993} and \citealp{Wachs2000} respectively.

\section{Visitor Distribution of Incoming Cars}
\begin{figure}[!ht]
	\begin{center}
	\includegraphics[width=140mm]{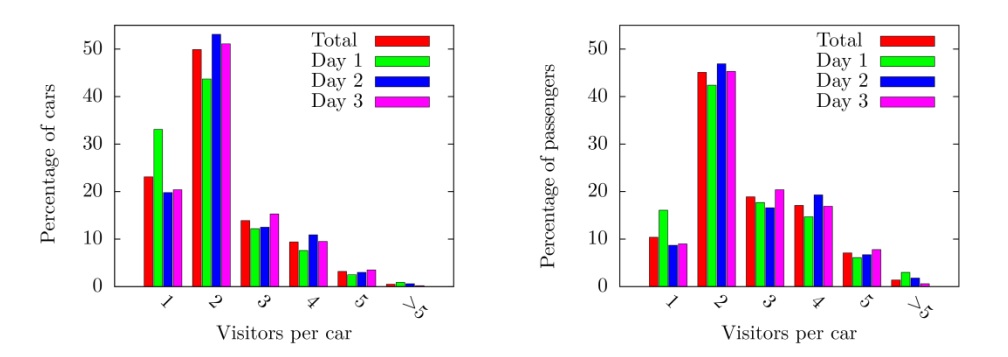}
	\caption{The Distribution According to The Total Number of Cars Counted and the Distribution According to The Total Number of Passengers}
	\label{figure2}
	\end{center}
\end{figure}
\noindent
To model a realistic transformation of pedestrians and cars, it is necessary to know the distribution of passengers per car. This means, that we have to know how many visitors are normally transported by one car. Therefore, we conducted an extensive field study to obtain the lacking data. We studied a large music festival in Munich over three consecutive days and counted the amount of visitors per car. We surveyed a total number of 1960 cars (for details see Table \ref{table:one}), which carried an average of 2.21 persons. Over 70\% of the vehicles were occupied by one or two passengers. A negligible amount of cars transported more than five people, and the maximum observed was eight passengers per car. The observed distributions from all three days can be seen in Figure \ref{figure2}. This data was used as a configuration for the transformation process of the hybrid model.
\begin{table}
\begin{center}
\caption{Number of Visitors Per Car on The Music Festival “Rockavaria”}
\label{table:one}
\begin{tabular}{|c|c|c|c|c|c|c|c|} \hline
Date & 1 & 2 & 3 & 4 & 5 & 5+ & Total \\\hline
29.05.2015 & 144 & 190 & 53 & 33 & 11 & 4 & 435 \\\hline
30.05.2015 & 98 & 263 & 62 & 54 & 15 & 3 & 495 \\\hline
31.05.2015 & 210 & 526 & 158 & 98 & 36 & 2 & 1030 \\\hline
\textbf{Total} & \textbf{452} & \textbf{979} & \textbf{273} & \textbf{185} & \textbf{62} & \textbf{9} & \textbf{1960} \\\hline
\end{tabular}
\end{center}
\end{table}

\section{The CarPed-Toolbox}

\subsection{Structure and Usability}
The software toolbox CarPed was developed as a proof of concept. Great attention was payed towards usability. Therefore, we implemented an intuitive and easy to use graphical user interface (GUI).\\\\
The GUI is divided into three major parts, namely the network input, the model and solving options and the result window. The network, consisting of nodes and edges, can be easily entered through the network input. All network depending properties, like the length or the capacity of an edge, can be determined immediately after the input procedure. By using the
model and solving options window, the user can determine which computational method should be used and how the data should be processed. Finally, the computed results can be accessed and visualized in the viewer. As all computed data is stored, the density distribution for each time-step can be looked at by the user. Different coloring of the edges shows the current densities. Alternatively, the user can click on an edge or node to get more detailed information about this object like the amount of pedestrians or cars on it. An exemplary screenshot of the GUI can be seen in Figure 3.
\begin{figure}[!ht]
	\begin{center}
	\includegraphics[width=140mm]{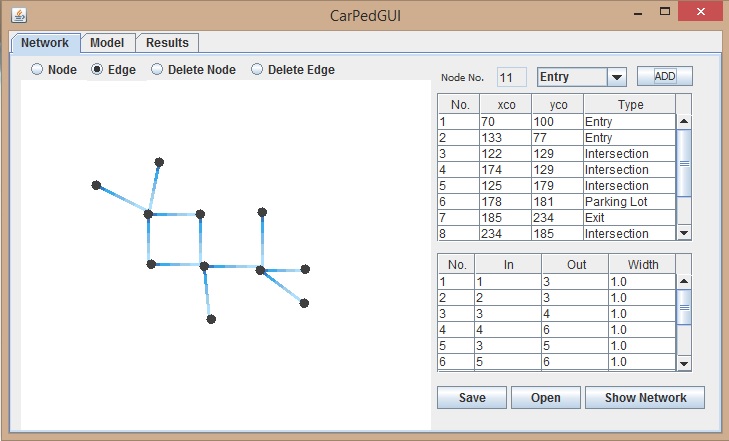}
	\caption{Representation of an Exemplary Simulation Scenario in CarPed}
	\label{figure3}
	\end{center}
\end{figure}

\subsection{Workflow}
The network can be entered by clicking on the designated area in the GUI. Special nodes, like entry and exit nodes or parking lots, can be defined by the user. Additionally, the nodes
can be linked to incoming or outgoing edges. If the input to the system is finished, a checking routine searches for errors and discrepancies. For each algorithm, the user can individually
modify the settings. After everything is set up, the solver can begin its calculations by just pressing a button.\\\\
At the beginning of the simulation, the network model reads the information of nodes and edges from the database previously created. It constructs a network from the infrastructure
and passes the entire network to the numerical solver. Together with the received input data, the numerical solver starts generating vehicles and pedestrians in the associated entry
nodes. It determines the initial density for each edge that is connected to the entry node and computes the density distribution. For each time-step, the results are passed from the numerical solver and stored in the network model. If a subject enters a node, it gets removed from this incoming edge and is put into an outgoing edge according to the route calculated by the Dijkstra algorithm. Additionally, the subject gets transformed if the visited node is a parking lot. The computations are repeated until all pedestrians and cars have reached an exit node and are dismissed from the system.\\\\
During the result analysis, the GUI displays two visions: a parameter set and an animation of density distributions. All statistical information is provided by the toolbox and can be
easily accessed.

\section{Conclusion and Outlook}
This report describes the dedicated hybrid and macroscopic simulation approach CarPed. This approach uses the macroscopic Hartmann and Sivers model on a network of nodes and edges. It is able to simulate human and vehicular flow in one hybrid model. The transformation between pedestrians and cars is carried out at special nodes, which represent parking lots.\\\\
The tool is thought to play a fundamental rule in the planning phase of massive public events. Organizers can use it to identify pathways and roads, which have a higher risk of congestions. Therefore, the organizers can develop strategies to avoid dangerous situations. In future studies, the current approach will be extended to a hybrid model for all-encompassing traffic control. This extended hybrid model will include more traffic subjects, like trains or ships to receive a global and universal traffic simulator.\\\\
Additionally, the CarPed simulator needs to be validated by field data. Therefore, we observed a public music festival over two consecutive years \citep{BiedermannRaspPi}. The first results are promising, but need further investigations.

\section{Acknowledgements}
We would like to thank our supervisors Daniel H. Biedermann and Peter K. Kielar for their nice suggestions, advices, and their unconditional help at any time. Also Isabella von Sivers for her support and long conversations according to this topic. Additionally, we would like to thank Andrea Mayer, Andreas Riedl and other student assistants for their help to collect sufficient data.\\

\section{Program Code}
\label{section:code}


\subsection{Subalgorithm for Traverse(edge)}

\begin{figure}
\centering\includegraphics[scale=0.5]{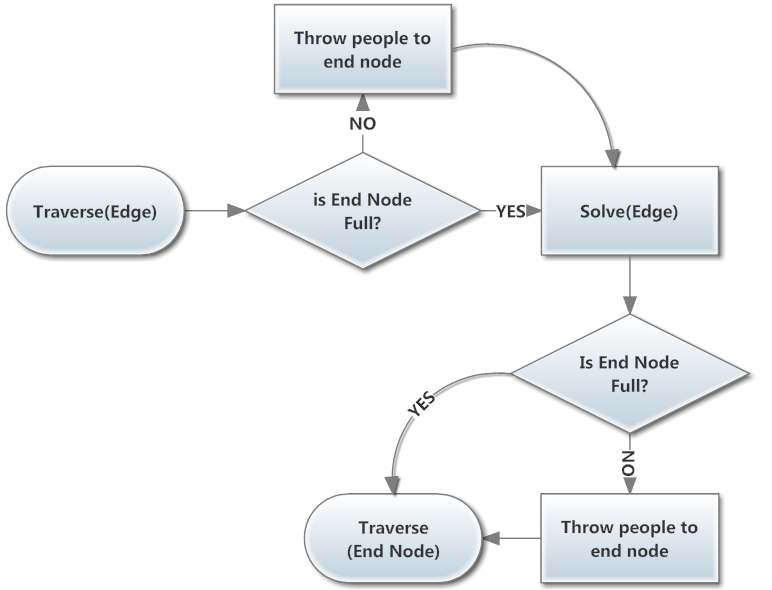}

\caption{Algorithm for Traverse(edge)}
\label{fig:traverseedge}

\end{figure}

The traverse algorithm in the edge first check if the end node is full, to see if there are still people in the penultimate cell. If that's the case, only solve the edge. If not throw the people in the penultimate cell into the node and solve the edge. The penultimate
available cell should be empty before the solver, otherwise mass is not conserved. That's why the solver (see figure \ref{fig:traverseedge}) has inside a Distribute Last Densities function that finds and empties the available penultimate cell. After solving, check again for full node to throw people. Finally, apply traverse to the end node.

\subsection{Subalgorithm Solve(Edge)}

\begin{figure}
\centering\includegraphics[scale=0.5]{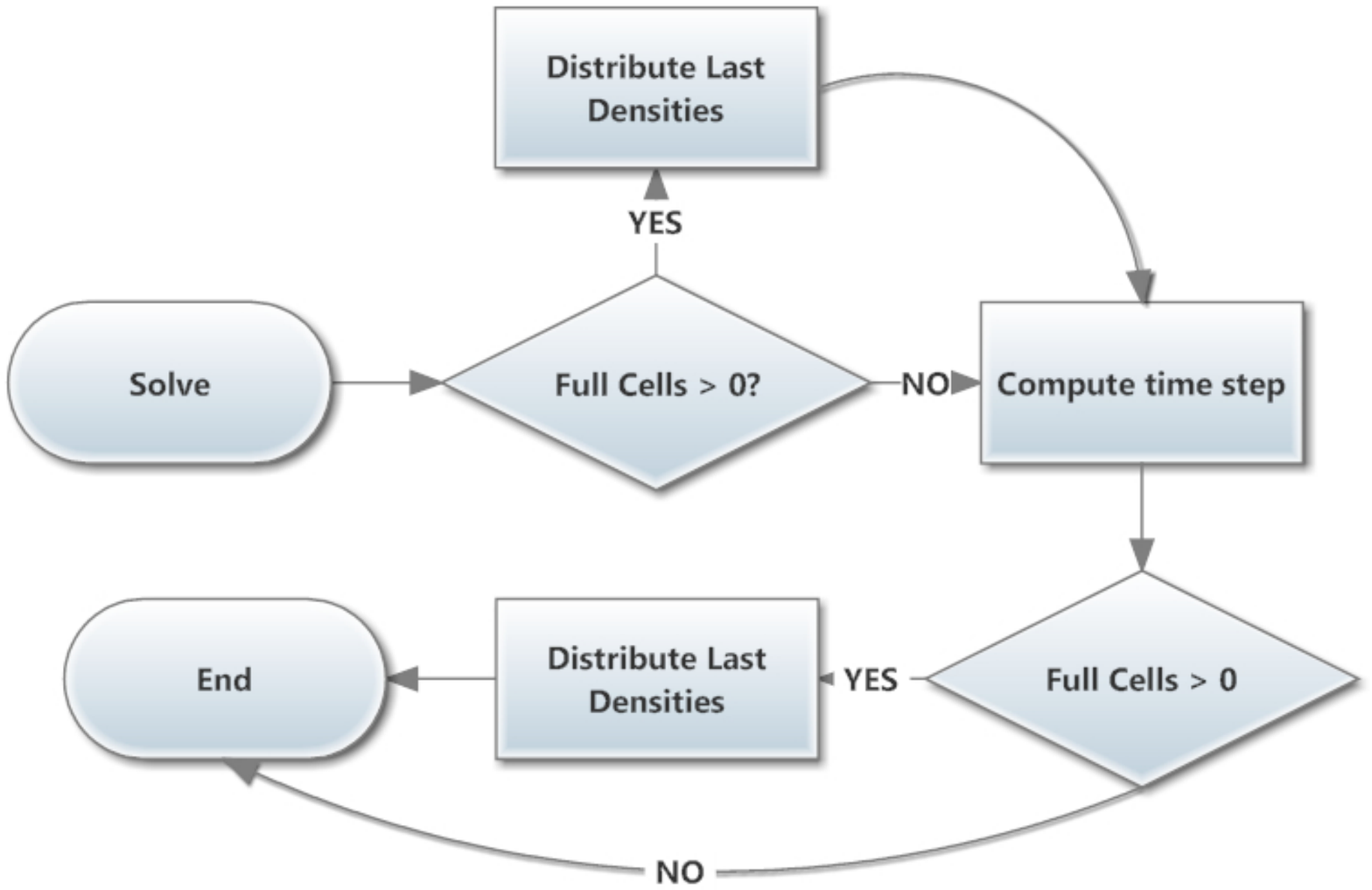}

\caption{Algorithm for Solve(Edge)}

\end{figure}

Check if there are full cells in edge. If thats the case, distribute last densities before and after computing the timestep.

\subsection{Sub algorithm for Traverse(Node)}

This traverse checks if Fixed or Dijkstra distributor is chosen.

\subsection{Sub algorithm for Distributor Fixed}

After all the incoming edges of the node threw people into it, we have a total amount to distribute into the outcoming ones. If that density to distribute cannot fit into the outcoming edges, the node is set full and it throws only the possible amount of people into
the outgoing edge, given by a friction which is nothing else that a quantity between 0 and 1, the percentage of people that fits. If density to distribute fits all, set the node not full and throw all the people into the outcoming edges.

\begin{figure}
\centering\includegraphics[scale=0.5]{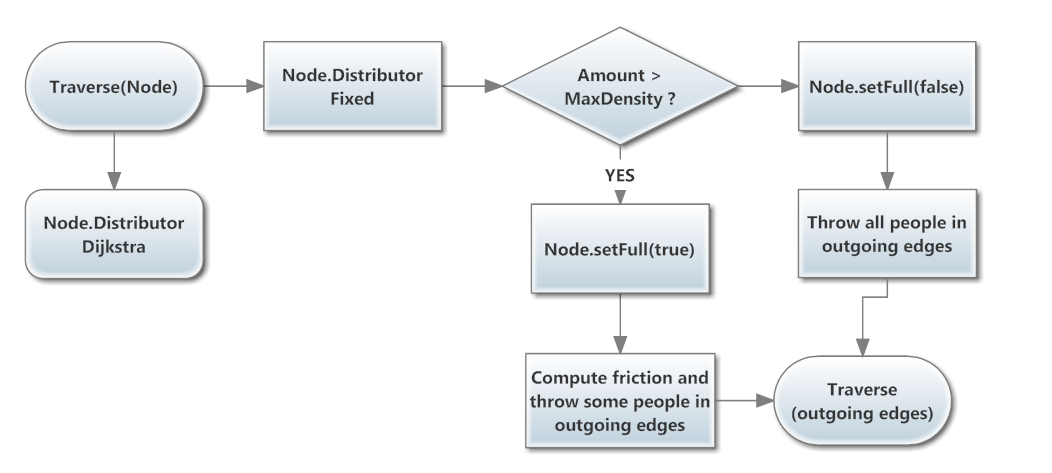}

\caption{Algorithm for Traverse(Node)}

\end{figure}

\subsection{Subalgorithm Node.SetFull(bool)}

If the input is true, check first if incoming edges have full cells. If that is the case do nothing, because there the Distribute Last Densities function in the edge already took the job of reorganizing the edge cells. If they have zero full cells, set one full cell for all of them. If false, set all full cells zero.

\begin{figure}
\centering\includegraphics[scale=0.5]{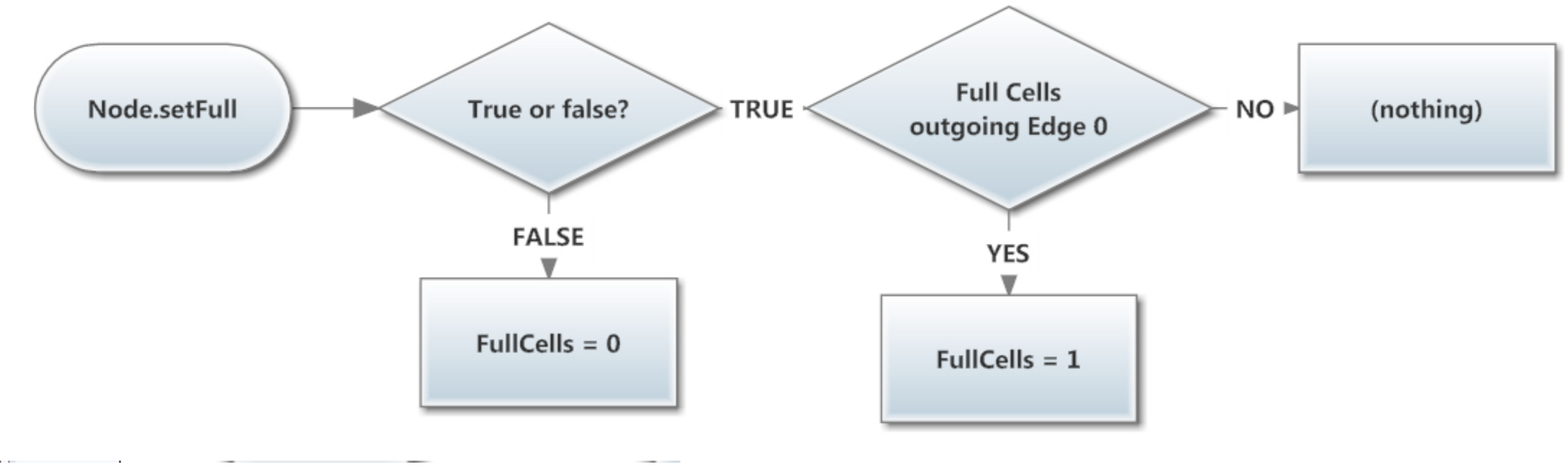}

\caption{Algorithm for Node.setFull(bool)}
\end{figure}

\subsection{Subalgorithm Edge.distributeLastDensities()}

The routine checks at first if both ultimate and penultimate denisties can be summed, using the amount. If this amount is more than the maximum density, define a friction and move only the possible people. After that, increase by one the number of full cells, and then call the
same function again recursively. If thats not the case, just move people from the penultimate cell to the ultimate one. 

\begin{figure}
\centering\includegraphics[scale=0.5]{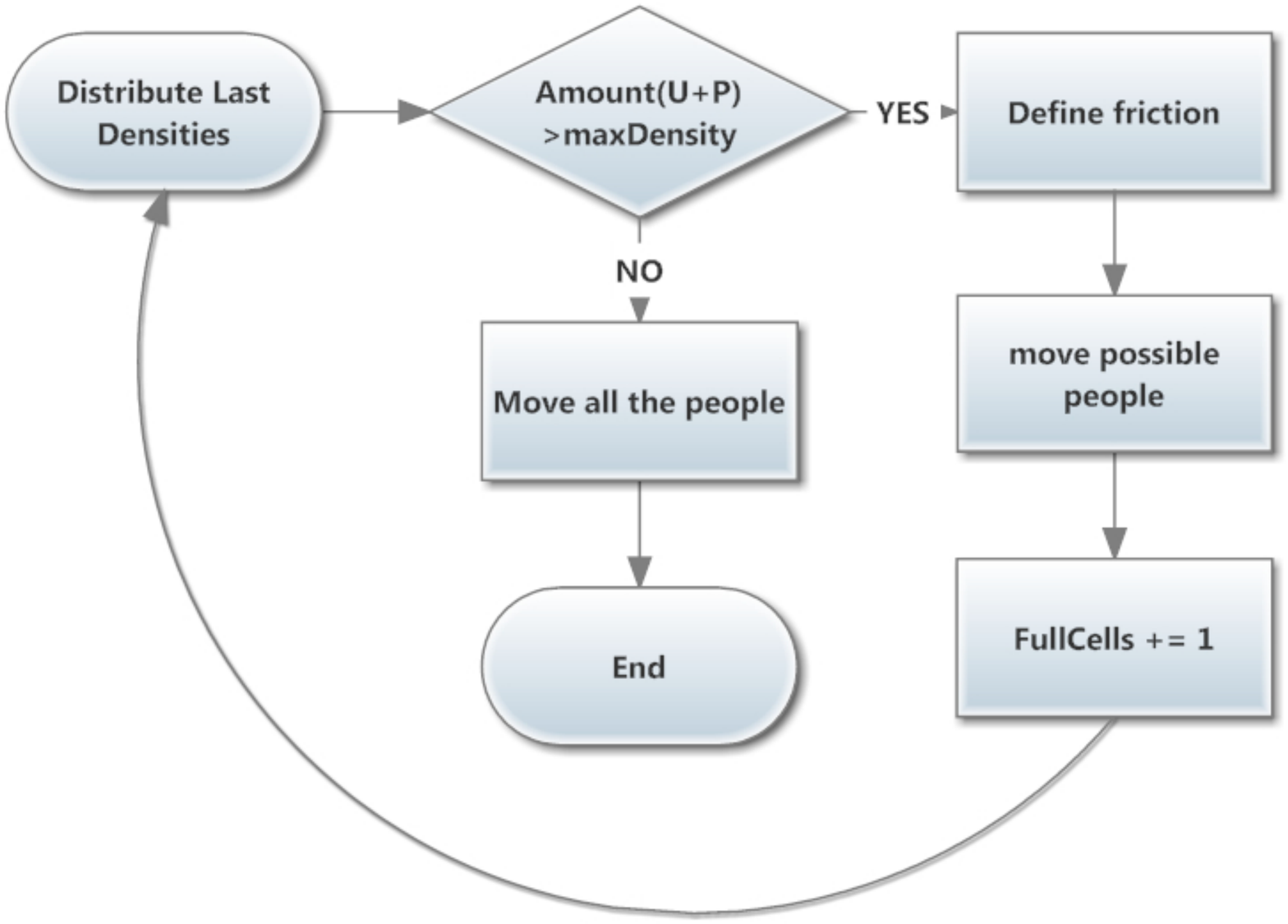}

\caption{Algorithm for Edge.distributeLastDensities()}

\end{figure}

\bibliography{ReferencesCarPed}

\end{document}